\documentclass[aps,prd,preprintnumbers,nofootinbib,showpacs,onecolumn,superscriptaddress]{revtex4}
\usepackage[dvips,dvipdfmx]{graphicx}
\usepackage{bm,latexsym,amsmath,amssymb,amsfonts,mathrsfs}
\usepackage{color}
\begin{document}

\newcommand{\simgt}{\lower.5ex\hbox{$\; \buildrel > \over \sim \;$}}
\newcommand{\simlt}{\lower.5ex\hbox{$\; \buildrel < \over \sim \;$}}

\newcommand{\nan}[1]{\textcolor{red}{#1}}
\newcommand{\yama}[1]{\textcolor{blue}{#1}}

\vspace{2cm}

\begin{flushright}
  HUPD-1811, SAGA-HE-293, KEK-TH-2095
  %{\tt hep-ph/08mmxxx}\\
\end{flushright}

\title{\large Large-scale inhomogeneity of dark energy produced in the ancestor vacuum}

\author{Yue Nan}
\affiliation{
  Department of Physics, Graduate School of Science, Hiroshima University,
  Higashi-Hiroshima 739-8526, Japan}

\author{Kazuhiro Yamamoto}
%\email{kazuhiro@hiroshima-u.ac.jp}
\affiliation{
  Department of Physics, Graduate School of Science, Hiroshima University,
  Higashi-Hiroshima 739-8526, Japan}
\affiliation{
  Department of Physics, Kyushu University, 744 Motooka, Nishi-Ku, Fukuoka 819-0395, Japan}

\author{Hajime Aoki}
\affiliation{
  Department of Physics, Saga University, Saga 840-8502, Japan}

\author{Satoshi Iso}
\affiliation{
Theory Center, High Energy Accelerator Research Organization (KEK),
and Graduate University for Advanced Studies (SOKENDAI), Ibaraki 305-0801, Japan}

\author{Daisuke Yamauchi}
\affiliation{
Faculty of Engineering, Kanagawa University, Kanagawa, 221-8686, Japan}

\begin{abstract}
  We investigate large-scale inhomogeneity of dark energy 
  in the bubble nucleation scenario of the universe.
  In this scenario, the present universe was created 
  by a bubble nucleation due to quantum tunneling 
  from a metastable ancestor vacuum, 
  followed by a primordial inflationary era. 
  During the bubble nucleation, 
  supercurvature modes of some kind of a scalar field are produced, 
  and remain until present without decaying; thus they can play a role of the
  dark energy, if the mass of the scalar field is 
  sufficiently light in the present universe.
  The supercurvature modes fluctuate at a very large spatial scale,
  much longer than the Hubble length in the present universe.
  Thus they create large-scale inhomogeneities of the dark energy,
  and generate large-scale anisotropies in the
  cosmic microwave background (CMB) fluctuations.
  This is a notable feature of this scenario,
  where quantum fluctuations of a scalar field
  are responsible for the dark energy.
  In this paper, 
  we calculate imprints of the scenario 
  on the CMB anisotropies through the integrated Sachs-Wolfe (ISW) effect,
  and give observational constraints on the curvature parameter $\Omega_K$ and on an
  additional parameter $\epsilon$ describing some properties of the ancestor vacuum.
\end{abstract}


\maketitle

\def\bar{\overline}
\def\x{{z}}
\def\L{L}
\def\bmx{{\bm x}}
\def\bmy{{\bm y}}
\def\bmp{{\bm p}}
\def\bmk{{\bm k}}
\def\deltatau{{\Delta\tau}}
\def\barphi{{\bar\phi}}
\def\barpi{{\pi_\chi}}

%%%%%%%%%%%%%%%%%%%%%%%%%%%%%%%%%%%%%%%%%%%%%%%%%%%%%%%%%%%%%%%%%%%%%%%%%%%%%%%%%%%%%%%%%%%%%%%%%%%%%%%%%%%%%%%%%%%%%%%%%%%%%%%%$
%%%%%%%%%%%%%%%%%%%%%%%%%%%%%%%%%%%%%%%%%%%%%%%%%%%%%%%%%%%%%%%%%%%%%%%%%%%%%%%%%%%%%%%%%%%%%%%%%%%%%%%%%%%%%%%%%%%%%%%%%%%%%%%%%

\section{Introduction}

The standard cosmological model, the $\Lambda\rm{CDM}$ model, describes the history 
of our universe which is composed of radiation, baryonic matter, cold dark matter (CDM) and 
dark energy represented by the cosmological constant $\Lambda$.
Observations of the large-scale structures have played important roles in 
determining the fraction of each component to the total energy density:
approximately 30\% for the matter components
and 70\% for the dark energy. 
Another possible ingredient, 
the spatial curvature of the universe $\Omega_K$, is known to be very close to zero, 
and our universe is almost spatially flat~\cite{Planck1,Planck2,Planck3,Planck2018}.
The $\Lambda\rm{CDM}$ model, together with the assumption of the primordial inflation, 
has successfully explained various cosmological observations: e.g., 
the cosmic microwave background (CMB) anisotropies, 
the abundance of light elements in the early universe, 
the baryonic acoustic oscillation (BAO) peaks, and the formation of the cosmological structures. 

Dark energy is the most dominant component of the present universe, and
accelerates the expansion of the universe~\cite{Weinberg}.
But its nature is unknown and the origin of the dark energy is the most intriguing riddle in the universe. 
The simplest hypothesis for dark energy is the cosmological constant $\Lambda$, 
which has survived various observational tests, by e.g.,
the Planck satellite, the Baryon Oscillation Spectroscopic Survey (BOSS) in the Sloan Digital Sky Survey (SDSS) project,
and the Dark Energy Survey (DES). 
On the other hand, 
it is recently argued that the cosmological constant cannot be compatible with string theory predictions
\cite{Obied,Agrawal,Ooguri,Krishnan}. 
Indeed, many dynamical scenarios for the dark energy predict 
different equations of state (EoS), $w \neq -1$, which can be tested or falsified by observations~\cite{ABM,TSJ,Vagnozzi:2018jhn}.
An example is the quintessence model based on a classically rolling scalar field \cite{Tsujikawa:2013fta}.
Other examples are based on quantum fluctuations of ultralight scalar fields (e.g. \cite{Ringeval,Glavan1,Glavan2,Glavan3,Glavan4,DEquantum,DEquantum2}). 
In this context, a connection with the string axiverse scenario is interesting~\cite{Arvanitaki,Witten,Visinelli:2018utg}. 

In the present paper, we investigate one of such dynamical scenarios of the dark energy~\cite{Aoki,Yamauchi}.
It is based on a bubble nucleation of our universe from a metastable {\it ancestor} vacuum in de Sitter spacetime.
The universe is assumed to be created by quantum tunneling of a scalar field, which is semi-classically described by
the Coleman-De Luccia (CDL) instanton~\cite{CDL}.
We note that bubble-nucleation transitions could be a
characteristic feature of the string landscape scenario \cite{FKML1,RHS,FKML2,GuthNomura}.
Following the scenario, the present universe is then described by an open Friedmann-Lemaitre-Robertson-Walker (FLRW)
universe with negative spatial curvature. 
In addition, we introduce a scalar field $\phi$, which is different from the CDL tunneling field. 
Then the tunneling from the ancestor vacuum generates very long wavelength modes of the $\phi$ field;
the supercurvature modes. These modes remain out of the horizon until present
and can play a role of the dark energy in the present epoch.  

The supercurvature modes are the so-called discrete modes and have an imaginary wave number
on the three-dimensional sphere $\rm{S^3}$ in the Euclidean CDL geometry.
Because of this, the modes are non-normalizable on the hyperbolic $\rm{H^3}$, 
when they are analytically continued from the Euclidean CDL geometry to the Lorentzian region
to describe a bubble nucleation in de Sitter spacetime~\cite{Sasaki,Yamamoto,Garr}. 
As long as the mass of the scalar field is sufficiently light, 
the supercurvature modes decay slowly at large distances, and give rise to long-range 
fluctuations of the field in the open universe. 
The length scale of the fluctuations, which is called the supercurvature scale $L_{sc}$, 
is much larger than the present spatial curvature scale $L_{c}$ of the Universe
and the Hubble length $H_0^{-1}$ at present;
$L_{sc} \gg L_{c} \gtrsim 10 H_0^{-1}$.
Thus, the supercurvature-mode energy density takes an almost
constant value within the horizon scale of the observable universe; it behaves as the dark energy, and
we call it the supercurvature-mode dark energy.
Possible observable signatures of the scenario in the EoS have been investigated
in Ref.~\cite{Yamauchi} with an expectation of being verified in the galaxy surveys by Square Kilometre
Array (SKA) and Euclid mission in the forthcoming decade~\cite{Amendola}. 
In the present paper, we further investigate another verifiable 
property of the supercurvature-mode dark energy. 
A novel feature of the supercurvature-mode dark energy is that the mode is not exactly
homogeneous and may induce tiny anisotropies and inhomogeneities of the dark energy
 even on the scale of the observable universe (cf. \cite{Mukhanov}). 
 The anisotropies are transformed into the anisotropic patterns of the CMB spectrum
 through the late-time integrated Sachs-Wolfe (ISW) effect, 
 which can distinguish the model from the simplest $\Lambda\rm{CDM}$ model.

The paper is organized as follows. 
In Sec.~\ref{sec:theo}, we will review the setup of the
supercurvature-mode dark energy scenario and calculate the spatial correlation of the dark
energy density contrast. 
In Sec.~\ref{sec:formal}, we calculate the two-point correlation function 
of the CMB fluctuations. The inhomogeneity of the supercurvature-mode dark energy 
is imprinted in the large-angle correlation of the CMB anisotropies.
Comparison with the observational data put upper bounds on the
curvature parameter $\Omega_{K}$ and the parameter $\epsilon$ that describes some properties
of the ancestor vacuum. 
Finally, in Sec.~\ref{sec:sum}, we will summarize the results.
Details of calculations are given in Appendices.

%%%%%%%%%%%%%%%%%%%%%%%%%%%%%%%%%%%%%%%%%%%%%%%%%%%%%%%%%%%%%%%%%%%%%%%%%%%%%%%%%%%%%%%%%%%%%%%%%%%%%%%%%%%%%%%%%%%%%%%%%%%%%%%%$
%%%%%%%%%%%%%%%%%%%%%%%%%%%%%%%%%%%%%%%%%%%%%%%%%%%%%%%%%%%%%%%%%%%%%%%%%%%%%%%%%%%%%%%%%%%%%%%%%%%%%%%%%%%%%%%%%%%%%%%%%%%%%%%%%

\section{Spatial correlation of the Supercurvature-mode dark energy}
\label{sec:theo}
In this section, 
we first briefly review the supercurvature-mode dark energy scenario following \cite{Aoki,Yamauchi}. 
In this model, the dark energy behaves nearly identical to the cosmological 
constant except for spatial inhomogeneities on very large scales.
Then, we calculate spatial variations of the dark energy, 
 which motivates the investigations of detectability through
 the CMB anisotropies in the next section. 
Suppose that our universe is an open universe
created by a bubble nucleation due to the CDL quantum tunneling of a scalar field~\cite{CDL}. 
After the bubble nucleation, the primordial inflation occurred first and then 
the big bang universe with negative spatial curvature has started. 
We also introduce another scalar field $\phi$ whose supercurvature mode is
generated through the bubble nucleation process. 
The mode will become the dark energy, which we call 
the supercurvature-mode dark energy. 
Before the tunneling, in the metastable de Sitter (ancestor) vacuum, 
Hubble parameter and mass of $\phi$ are denoted by $H_A$ and $m_A$, respectively.
Ordinary inflation follows the bubble nucleation in the hyperbolic spatial
geometry. The Hubble parameter of the inflation is denoted by $H_I$.
We note that the Hubble parameters before and after the CDL quantum tunneling
satisfy the relation $H_A>H_I$~\cite{Aoki}. The mass of the
scalar field $\phi$ after the tunneling is set $m_0$, which could be different
from $m_A$. 

In the free field approximation,
we can solve the equation of motion for the scalar field $\phi$
on the CDL background in Euclidean space;
expanding solutions in terms of the eigenfunctions on 
 the 3-dimensional sphere slice ${\rm S}^3$ with eigenvalues $-(k^2+1)$, 
the equation of motion becomes a 
Schr\"{o}dinger-like equation with a finite potential.
The eigenfunctions on ${\rm S}^3$ are classified into two types of modes.
One type is a continuous mode with a real wave number $k$ while the other 
is a discrete mode with an imaginary wave number $k=i(1-\epsilon)$. 
The discrete mode is called the supercurvature mode. 
Here $\epsilon$ is determined by the properties of the ancestor vacuum
and given by 
\begin{eqnarray}
 \epsilon=c_\epsilon\left(\frac{m_A}{H_A}\right)^2,
 \label{epsil}
\end{eqnarray}
where $c_\epsilon$ is an order $\mathcal{O}(1)$ quantity that depends on the critical size of
the bubble created in the ancestor vacuum.
The mass $m_A$ of the scalar field 
and the Hubble parameter $H_A$ in the ancestor vacuum are assumed 
to obey the condition $m_A \ll H_A$; thus $\epsilon \ll 1$ follows.
Analytically continued to the Lorentzian region in de Sitter space, 
the supercurvature mode becomes 
non-normalizable on the spatial slicing $\rm{H^3}$ of the open universe
and generate large-scale fluctuations. 
Unlike the continuous modes that decay as $\rm{e}^{-\eta}$
in the conformal time $\eta$, the discrete supercurvature mode behaves
$\rm{e}^{-\epsilon \eta}$ and decay remarkably slowly compared with the continuous modes. 
The scalar field is assumed to have ultralight mass $m_0<H_0\sim 10^{-33}\rm{eV}$, 
and the supercurvature mode plays a role of the dark energy in the present universe. 
A candidate of such ultralight fields may appear as an
axion-like particle (ALP) in string theory~\cite{Arvanitaki,Witten}. 
%In a recent work, properties of ALPs associated with the string axiverse to play cosmological roles 
%are constrained from cosmological data~\cite{Visinelli:2018utg}.

\begin{figure}[t]
  \begin{center}
    \includegraphics[width=90mm]{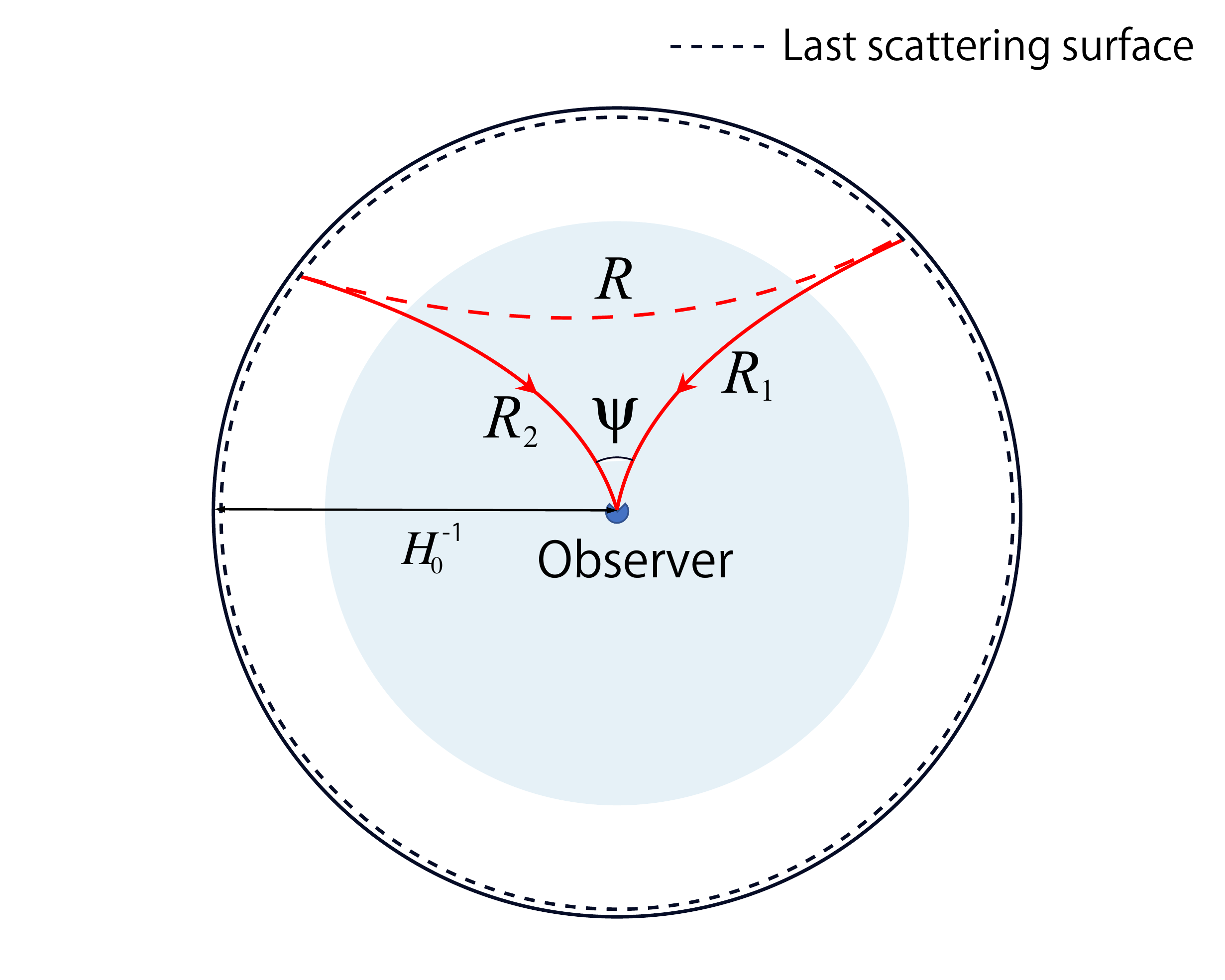}
  \end{center}
      \vspace{-0.cm}
\caption{Schematic of the choice of $\{R, R_1, R_2\}$ triplets along different line-of-sight
  for the two-point correlation function.}
\label{fig:schematic}
\end{figure}

In the following, we focus on the supercurvature modes and investigate
its properties as the dark energy in the present universe. 
The supercurvature mode contributes to the correlation function of the scalar field $\phi(x)$
in the open universe within the bubble as \cite{Aoki}
\begin{eqnarray}
  \langle\phi(\eta,\bm x)\phi(\eta',\bm x')\rangle
=\varphi(\eta)\varphi(\eta') 
{\sinh(1-\epsilon)R
  \over(1-\epsilon)\sinh R},
\label{2pcfphi}
\end{eqnarray}
where $\eta$ is the conformal time, $\varphi(\eta)$ is the frozen expectation value of field $\phi$. %(see
The explicit form of $\varphi$ is given in Eq.~(\ref{appen:varphi}).
%).
$R$ is the (dimensionless) geodesic distance 
on  the three-dimensional hyperbolic space ${\rm H}^3$,  
normalized in terms of the curvature scale $L_c=1/\sqrt{-K}$, 
and is given by
\begin{eqnarray}
  \cosh R&=&\cosh R_1\cosh R_2-\sinh R_1\sinh R_2\cos\psi.
  \label{RRRR}
\end{eqnarray}
$R_1$ and $R_2$ are the radial coordinates of 
the two points, $\bm x$ and $\bm x'$, and $\psi$ is
the included angle between them in the three-dimensional space (see Fig.~\ref{fig:schematic}). 
In the next section, we also use $\chi$ to denote the comoving radial coordinate distance with
dimension of the length; $R=\sqrt{-K}\chi$. Thus the curvature radius is given by $R_c=\sqrt{-K} L_c =1.$
%%%%%%%%%%%%%%%%%%%%%%%%%%%%%
%%%%%%%%%%%%%%%%%%%%%%%%%%%%

\begin{figure}[t]
\includegraphics[width=0.45\textwidth]{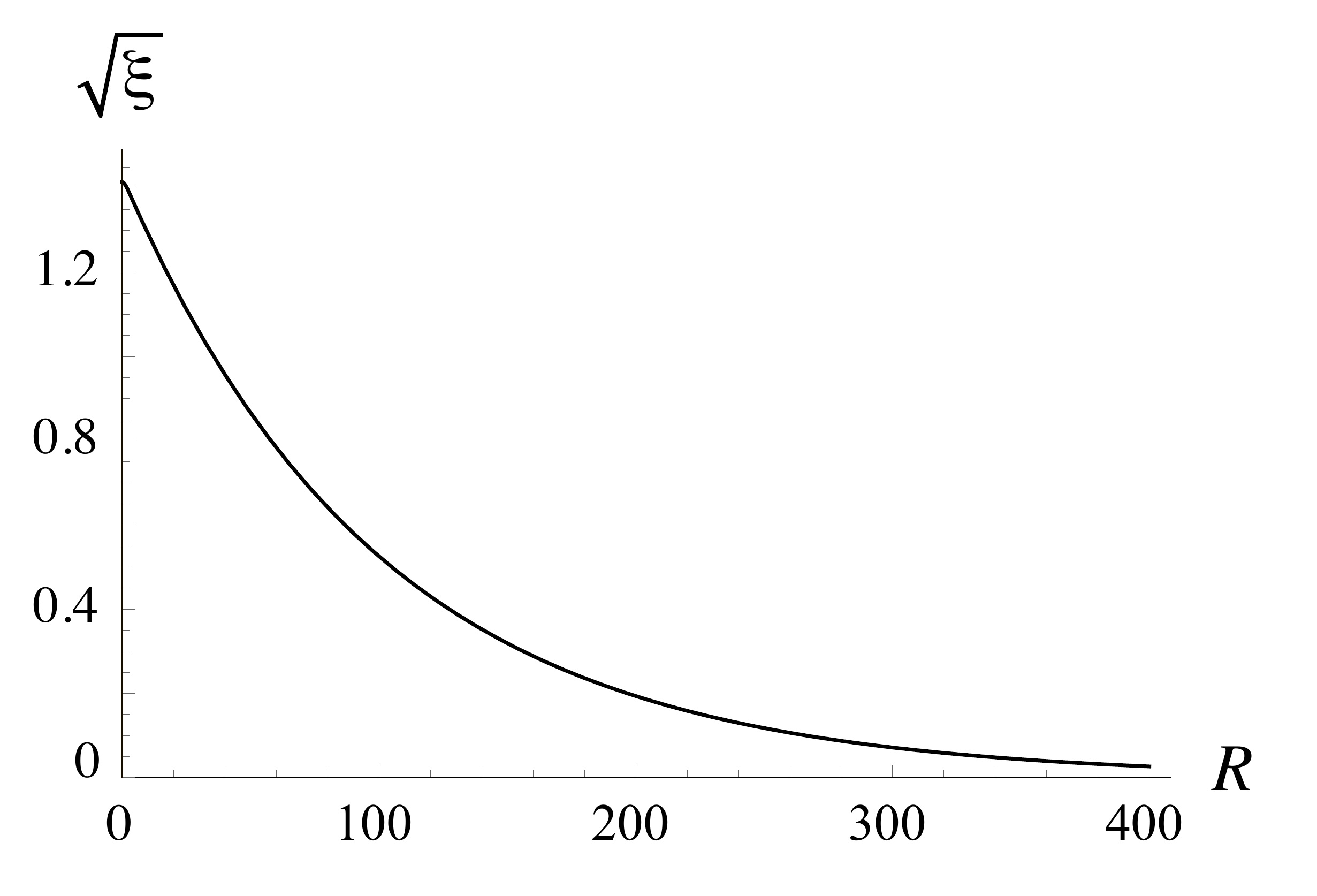} %eps}
\caption{
  $\xi^{1/2}(R)$ as a function of $R$, where we adopted $\epsilon=0.01$.
  The horizon scale at the present epoch is $R\sim\sqrt{-K}/H_0=\sqrt{\Omega_K} \ll 1$, 
  the curvature scale is $R=1$, and the supercurvature scale is $R=1/\epsilon \gg 1$.
  \label{fig:dSflat}}
\end{figure}

\begin{figure}[b]
    \includegraphics[width=80mm]{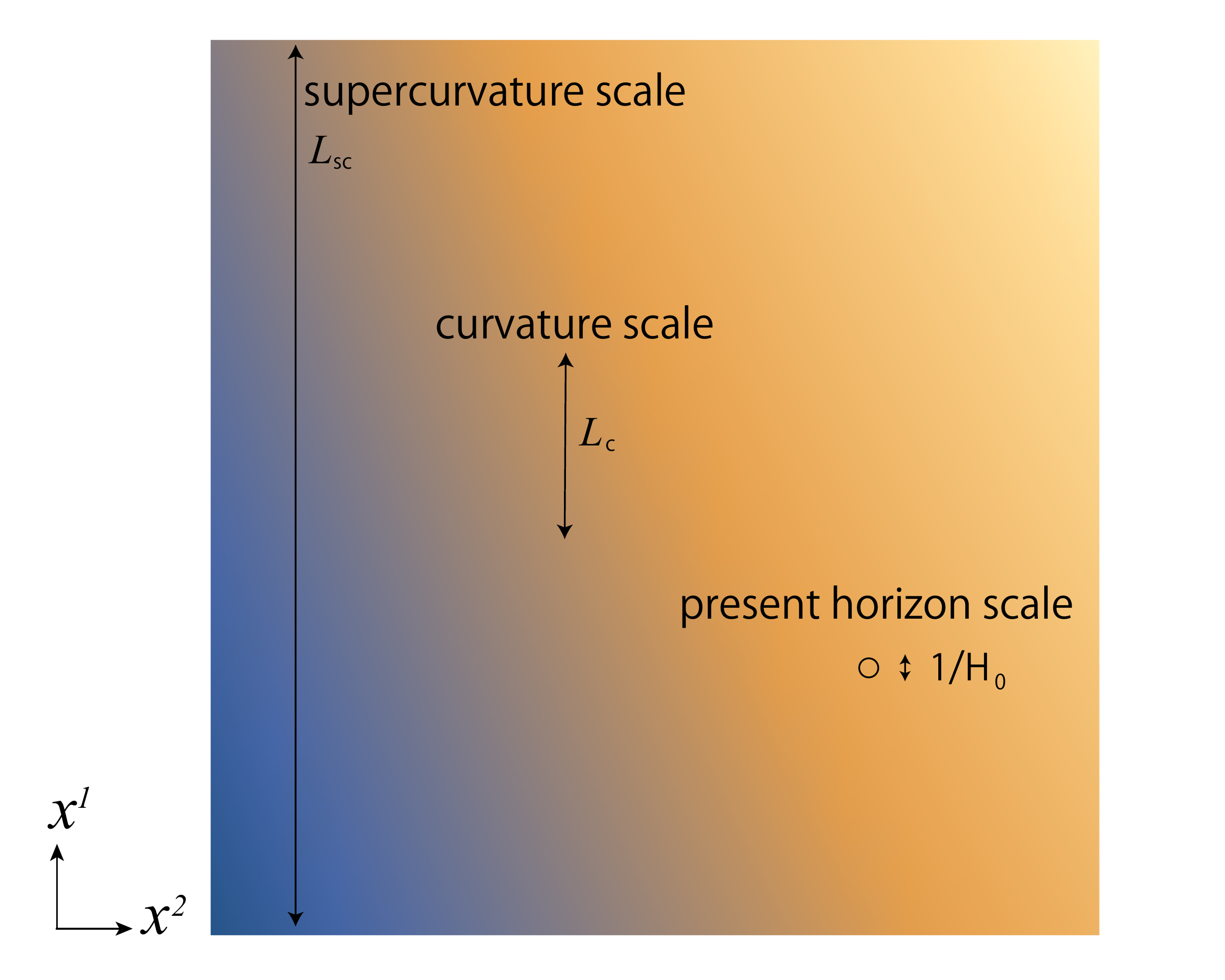}
  \caption{
    Schematic for supercurvature-mode dark energy density contrast,
    where the brightness of the color denotes the relative amplitude of the density contrast.
    We assume that the supercurvature scale $L_{sc}=1/\epsilon\sqrt{-K}(=L_c/\epsilon)$ is far beyond the curvature scale $L_c=1/\sqrt{-K}$.
    The curvature scale $L_c$ is beyond the comoving horizon scale so that the 
    observable universe appears flat. The horizon scale at the present epoch is $1/H_0$.
    Thus, we assume ${1/H_0\ll L_c\ll L_{sc}}$. 
    \label{fig:scales}}
\end{figure}

%%%%%%%%%%%%%%%%%%%%%%%%%%%%%%%%%%%%%%%%%%%%%%%%%%%%%%%%%%%%%%%%%%%%%%%%%%%%%%%%%%%%%%%%%%%%%%%%%%%%%%%%%%%%%%%%%%%
Now we calculate the spatial variation of the supercurvature mode dark energy.
In this model, the dark energy density at late times in the matter-dominated era 
(i.e., in the period (ii) in Sec. V-C of Ref.~\cite{Aoki}) is
dominantly given by the mass term in the energy-momentum tensor of the supercurvature mode\footnote{
At earlier times (i.e. in the period (i) in Sec. V-C of Ref.~\cite{Aoki}),
the spatial-derivative term dominates over the mass term
and the EoS parameter $w$ approaches $-1/3$.
Possible observational signatures of the time-dependent EoS
are investigated in Ref.~\cite{Yamauchi}.
However, for the analyses of the large-scale inhomogeneity that will be studied in
the present paper, this effect only gives higher order corrections,
and we will ignore it in the followings.
See Appendix~\ref{appen:eos} for detailed explanation.},
\begin{eqnarray}
  &&   \rho_{\rm DE}(\eta,\bm x)\simeq{m_0^2\over 2} \phi^2(\eta,\bm x) .
\end{eqnarray}
Then defining  the density contrast of the dark energy by
\begin{eqnarray}
  &&   \delta(\eta,\bm x)={\rho_{\rm DE}(\eta,\bm x)-\langle\rho_{\rm DE}(\eta,\bm x)\rangle \over
    \langle\rho_{\rm DE}(\eta,\bm x)\rangle}
  \simeq {\phi^2(\eta,\bm x)-\langle \phi^2(\eta,\bm x)\rangle\over \langle \phi^2(\eta,\bm x)\rangle} ,
\end{eqnarray} 
 two point function of the density contrast can be calculated as
\begin{eqnarray}
    \langle\delta(\eta,\bm x)\delta(\eta,\bm y)\rangle
 &=&{\langle\phi^2(\eta,\bm x)\phi^2(\eta,\bm y)\rangle-
\langle \phi^2(\eta,\bm x)\rangle\langle \phi^2(\eta,\bm y)\rangle
   \over \langle \phi^2(\eta,\bm x)\rangle^2},
\end{eqnarray}
where we used 
$\langle \phi^2(\eta,\bm x)\rangle=\langle \phi^2(\eta,\bm y)\rangle$.
Furthermore, in the free field approximation, we can decompose the four-point function of $\phi$
into a product of two-point functions by using the Wick theorem:
\begin{eqnarray}
  \langle\phi^2(\eta,\bm x)\phi^2(\eta,\bm y)\rangle=
  \langle\phi^2(\eta,\bm x)\rangle\langle\phi^2(\eta,\bm y)\rangle
  +2\langle\phi(\eta,\bm x)\phi(\eta,\bm y)\rangle^2.
  \label{p2p2}
\end{eqnarray} 
Then, using Eq.~(\ref{2pcfphi}),   we have
\begin{eqnarray}
 \xi(R) \equiv  \langle\delta(\eta,\bm x)\delta(\eta,\bm y)\rangle ={2\langle\phi(\eta,\bm x)\phi(\eta,\bm y)\rangle^2
    \over\langle \phi^2(\eta,\bm x)\rangle^2}
  = 2\left({\sinh(1-\epsilon)R\over(1-\epsilon)
    \sinh R}\right)^2 ,%\sim e^{-2\epsilon R}
\end{eqnarray}
where $R=\sqrt{-K}|\bm x-\bm y|$.
The correlation function $\xi(R)$ changes its behavior around the curvature scale $R_c=1$ as
\begin{eqnarray}
  \xi(R)
  \simeq
  2\times\left\{
  \begin{array}{cc}
    1& ~~ R\ll 1 \\
    e^{-2\epsilon R}& ~~ R\gg 1 \\
   \end{array}
\right. ,
\end{eqnarray}
and diminishes at distances over the supercurvature scale $R_{sc} \equiv1/\epsilon$. 
In physical length, $R_{sc}$ corresponds to $L_{sc}=L_c/\epsilon$,
which is much larger than the curvature radius $L_c$. 
The behavior of $\sqrt{\xi(R)}$ for $R \gg R_{sc}$ is depicted in Fig.~\ref{fig:dSflat}. 
This indicates that the supercurvature-mode dark energy density 
varies considerably beyond the supercurvature scale $R_{sc}$.
In Fig.~\ref{fig:scales}, we show a schematic
picture of the spatial variation of the supercurvature-mode dark energy. 
At the horizon scale $H_0^{-1} (\ll L_c)$, we 
take $R=\sqrt{-K} H_0^{-1}=\sqrt{\Omega_K}$, 
where we used the relation $\Omega_K=-K/H_0^2$. 
For $\Omega_K,\epsilon \ll 1$, we have 
\begin{eqnarray}
\sqrt{ \langle\delta^2(0)\rangle }- \sqrt{ \langle\delta(0)\delta(1/H_0)\rangle }
=\sqrt{2}-\sqrt{2}{\sinh(1-\epsilon)\sqrt{\Omega_K}\over(1-\epsilon)\sinh \sqrt{\Omega_K}}
\simeq\sqrt{2}{\epsilon\Omega_K\over 3} ,
\label{Deltadelta}
\end{eqnarray} 
which is extremely tiny ($\propto \epsilon \Omega_K$). However, as we will see
in the next section, 
it may give rise to an observable effect in the CMB anisotropies
on the large scales. 

%%%%%%%%%%%%%%%%%%%%%%%%%%%%%%%%%%%%%%%%%%%%%%%%%%%%%%%%%%%%%%%%%%%%%%%%%%%%%%%%%%%%%%%%%%%%%%
%%%%%%%%%%%%%%%%%%%%%%%%%%%%%%%%%%%%%%%%%%%%%%%%%%%%%%%%%%%%%%%%%%%%%%%%%%%%%%%%%%%%%%%%%%%%%%
\vspace{0mm}
The above results show that the density contrast of the supercurvature-mode dark energy
has an inhomogeneity of the order one over the scales of the supercurvature
$R \gtrsim R_{sc} \gg 1.$
 This large-scale variation of the dark energy density is the characteristic feature
 of the dark energy model based on quantum fluctuations.
 For the large scales $R>R_{sc}$, the dark energy density largely fluctuates and
 can be treated as a classical Gaussian random variable
 with the properties of $\langle \phi_{sc} \rangle=0$ and $\langle \phi_{sc}^2 \rangle=\varphi^2(\eta)$ (See Appendix \ref{appen:scmde} for the explicit expression of $\varphi(\eta)$.) 
 On the other hand, the dark energy density is nearly constant within
 the horizon $H_0^{-1} {(\ll L_c)}$. The explicit form of the probability distribution function
 of the dark energy density is shown in Appendix \ref{appen:pdf2}. 
 The result demonstrates a wide distribution of probability of $\rho_{\rm DE}$
 and the dark energy density parameter $\Omega_\Lambda$ at scales larger than
 the supercurvature scale $R_{sc}$ even when we fix the parameter
\begin{eqnarray}
 \langle\rho_{\rm DE}( \bm x)\rangle=
{1\over 2}m_0^2\varphi^2(0) =3H_0^2\bar\Omega_\Lambda
/ 8\pi G
\label{expectationrhode}
\end{eqnarray}
with $\bar\Omega_\Lambda=0.7$.
Thus the dark energy density has a large spatial variation on the large scales
$R>R_{sc}$. We also note that even within the horizon scale, $H_0^{-1}$,
there exists the spatial variation, though it is tiny as Eq.~(\ref{Deltadelta}).
In the next section, we will study the CMB anisotropies caused by it,
and give observational constraints on the model parameters of the scenario.

%%%%%%%%%%%%%%%%%%%%%%%%%%%%%%%%%%%%%%%%%%%%%
%%%%%%%%%%%%%%%%%%%%%%%%%%%%%%%%%%%%%%%%%%%%%
%%%%%%%%%%%%%%%%%%%%%%%%%%%%%%%%%%%%%%%%%%%%%
%%%%%%%%%%%%%%%%%%%%%%%%%%%%%%%%%%%%%%%%%%%%%
\section{CMB anisotropies from the supercurvature-mode}
\label{sec:formal}
To study observable effects from the spatial variations of the supercurvature-mode dark energy, we investigate  possible imprints from the supercurvature-mode
dark energy on the CMB anisotropies through the late-time ISW 
effect.
We adopt the line element under the conformal Newtonian gauge as
\begin{eqnarray}
  ds^2=a^2(\eta)\left[-(1+2\Psi)d\eta^2+(1+2\Phi)\gamma_{ij}dx^i dx^j\right],
\end{eqnarray}	
where 
$\Psi$ and $\Phi$ are the gravitational potential
and the curvature potential, respectively, and $\gamma_{ij}$ is the three-dimensional
metric in an open universe,
\begin{eqnarray}
    \gamma_{ij}dx^i dx^j=d\chi^2+\left({\sinh\sqrt{-K}\chi}\over \sqrt{-K}\right)^2
    (d\theta^2+\sin^2\theta d\varphi^2).
\end{eqnarray}

The evolution of the distribution function of CMB photons is
described by the Boltzmann equation with the perturbed Planck distribution: 
\begin{eqnarray}
 f(\eta, \bm x, \bm p)={1\over \exp[p/(T(\eta)(1+\Theta (\eta,\bm x,\bm \gamma) )]-1},
\end{eqnarray}
where $\Theta(\eta,\bm x,\bm \gamma)$ denotes 
 the temperature fluctuation of photons. ${\bm \gamma}$ is
the line-of-sight direction identical to the unit vector of the observed photon momentum
$\bm p$, while $p$ is its magnitude.
Note that the temperature fluctuation $\Theta(\eta,\bm x,\bm \gamma)$ 
depends on the photon's trajectory scattered from the past. 
It can be shown that the CMB anisotropy $\Theta(\eta,\bm x,\bm \gamma)$ satisfies
the equation~\cite{WHuthesis}
\begin{eqnarray}
  {d\over d\eta} (\Theta+\Psi)={\partial \Psi(\eta,\bm x) \over \partial \eta}-
  {\partial \Phi(\eta,\bm x) \over \partial \eta}+C_{\rm{e}\gamma},
\end{eqnarray}
where $C_{e\gamma}$ denotes the collision term for the Compton scattering,
but it can be omitted in our investigation.
Then, the integration yields the ISW contribution to the CMB anisotropies,
\begin{eqnarray}
 {\Delta T\over T}(\bm \gamma)=  \Theta(\eta_0,\bm x_0, \bm \gamma)+\Psi(\eta_0,\bm x_0)
  =\int_{\eta_*}^{\eta_0} d\eta\left(
    {\partial \Psi(\eta,\chi,\bm \gamma) \over \partial \eta}-
    {\partial \Phi(\eta,\chi,\bm \gamma) \over \partial \eta}\right)\biggl|_{\chi=\eta_0-\eta}.
\label{abc}
\end{eqnarray}
Here, on the right-hand side, the spatial position ${\bm x}$ is represented by 
its radial coordinate and the angle as ${\bm x}=(\chi, {\bm\gamma})$.
The direction of the photon, ${\bm\gamma}$, is fixed in this expression 
and the radial coordinate $\chi = \eta_0 -\eta$ 
denotes the position of the photon at the conformal time $\eta$. 
$\eta_*$ stands for the conformal time of the CMB last scattering surface. 
Hereafter, we use {\it dot} to denote a differentiation with respect to the conformal time
$\eta$, $\dot{} \equiv {\partial/ \partial\eta}$. 
In the following, we calculate the right-hand-side of Eq.~(\ref{abc}) by using 
 the perturbed Einstein equation. 

Using an overbar to represent the background quantity, 
we have the 0th order Einstein equation~\cite{WHuthesis}, 
\begin{eqnarray}
  \bar G^0{}_0=  {-3}{1\over a^2}\left({\cal H}^2+K\right)=8\pi G\bar T^0{}_0
  =8\pi G\left(\bar T^0{}_{0({\rm m})}+\bar T^0{}_{0(\phi)}\right),
\end{eqnarray}
where (m) and ($\phi$) denotes the matter component and the dark energy component,
respectively, and we defined ${\cal H}= \dot a/a=a_{,\eta}/a$.
A definition of the overbar will be shown momentarily.
The 1st order perturbation of the Einstein equation is given by 
\begin{eqnarray}
  \delta G^0{}_0=2{1\over a^2}\left[ 3{\cal H}^2\Psi-3{\cal H} \dot\Phi +(\nabla_H^2+3K)\Phi\right]
=  8\pi G\delta T^0{}_0=8\pi G\left(\delta T^0{}_{0({\rm m})}+\delta T^0{}_{0(\phi)}\right),
\label{dtt}
\end{eqnarray}
where 
$\nabla_H^2$ is the Laplacian defined with respect to $\gamma_{ij}$ as
$\nabla_H^2Q=\gamma^{ij} Q_{|ij}$ (See e.g., Ref. \cite{WHuthesis}).

On the other hand, the energy-momentum tensor for the scalar field $\phi$ is given by 
\begin{eqnarray}
  T^0{}_{0(\phi)}=-{1\over 2a^2}\left((1-2\Psi)\dot\phi^2+(1-2\Phi)\gamma^{ij}\nabla_i\phi\nabla_j\phi+{m_0^2a^2\phi^2}  \right).
\label{EMT00}
\end{eqnarray}
Its spatial average surrounding our horizon is defined as
\begin{eqnarray}
  &&  \bar T^0{}_{0(\phi)}=-{1\over 2a^2}\left((1-2\Psi)\dot\phi^2+(1-2\Phi)\gamma^{ij}\nabla_i\phi\nabla_j\phi+{m_0^2a^2\phi^2}  \right)\Bigr|_{{\rm SA} \chi=0},
\end{eqnarray}
where ``$\rm{SA}$'' denotes ``spatially average around'' surrounding the present Hubble scale of our Universe.
We then consider the fluctuation of $T^0{}_{0(\phi)}$ around $\bar T^0{}_{0(\phi)}$,
\begin{eqnarray}
  && \delta T^0{}_{0(\phi)} \equiv T^0{}_{0(\phi)}-\bar T^0{}_{0(\phi)} .
\end{eqnarray}
Since we are interested in the supercurvature-mode dark energy which
is almost constant within the Hubble scale, 
we approximate the spatially averaged value by the quantity at the observer $\chi=0$.
For example, we have 
\begin{equation}
\phi(\eta,\chi,\bm\gamma)|_{{\rm SA} \chi=0}=\phi(\eta,\chi=0,\bm\gamma).
\end{equation}
Of course, $\phi(\eta,\chi=0,\bm\gamma)$ does not depend on the direction $\bm\gamma$ and
we can simply write it as $\phi(\eta,\chi=0)$. 
As we are interested in the dark energy component that fluctuates mildly both 
in space and time, the mass term in the energy-momentum tensor (\ref{EMT00}) dominantly contributes:
$ T^0{}_{0(\phi)}\simeq -{1\over 2}{m_0^2\phi^2}$.
Then the background and the spatial fluctuation of $T^0{}_0$ are given by 
\begin{eqnarray}
\bar{T}^0{}_{0(\phi)}=-{1\over 2}{m_0^2\phi^2}\Bigr|_{{\rm SA}\chi=0} \ , \hspace{5mm}
 \delta T^0{}_{0(\phi)}=-{1\over 2}{m_0^2}
 \left( \phi^2 -  \phi^2 \Bigr|_{{\rm SA} \chi=0}\right) ,
  \label{deltaT00}
\end{eqnarray}
respectively. 

\vspace{2mm}
Now let us calculate the temperature fluctuation induced by the autocorrelation of the 
supercurvature-mode dark energy.
Since we are interested in the perturbations on the supercurvature scales, see Fig.~\ref{fig:scales},
the metric perturbation in the late-time universe can be approximated as $\Psi+\Phi=0$ and
the term $(\nabla_H^2+3K)\Phi$ is negligibly small. 
This allows us to approximate Eq.~(\ref{dtt}) as
\begin{eqnarray}
 \delta G^0{}_0=2{1\over a^2}\left[ 3{\cal H}^2\Psi+3{\cal H} \dot\Psi \right]
 =8\pi G\bigl(\delta T^0{}_{0(\phi)}
 +\delta T^0{}_{0({\rm m})}\bigr).
\label{dtt2}
\end{eqnarray}
The perturbed energy momentum tensor of the matter component is
$\delta T^0{}_{0({\rm m})}=-\delta_{\rm m}\rho_{\rm m}$, where $\delta_{\rm m}$
is the density contrast of the matter component, 
which follows (e.g., \cite{WHuthesis})
\begin{eqnarray}
  && \dot \delta_{\rm m}+kV_{\rm m}+3\dot \Phi=0,
  \\
  && \dot V_{\rm m}+{\dot a\over a}V_{\rm m}-k\Psi=0.
\end{eqnarray}
{
Here we follow the notation of Ref.~\cite{WHuthesis} for the Fourier expansion
in an open universe. Therefore, it should be understood that $k^2=-K(2\epsilon-\epsilon^2)$
for the supercurvature mode.}
These equations yield
\begin{eqnarray}
   (a \dot \delta_{\rm m})\dot{}+k^2a\Psi+3(a\dot\Phi)\dot{}=0,
\end{eqnarray}
where we may omit the term of the gravitational potential $k^2a\Psi$,
in the limit of the large scales, as we consider the supercurvature mode.
Then, we have
\begin{eqnarray}
   \delta_{\rm m}(\eta)+3\Phi(\eta)=0,
\end{eqnarray}
where we assumed $\delta_{\rm m}(0)=\Phi(0)=0$ for the supercurvature mode
perturbations. With $\Psi+\Phi=0$, Eq.~(\ref{dtt2}) reduces to
\begin{eqnarray}
  6{{\cal H}\over a^2}\dot \Psi+\left(6{{\cal H}^2\over a^2}+24\pi G\rho_m\right)
  \Psi=8\pi G \delta T^0{}_{0(\phi)}.
\label{dtt22}
\end{eqnarray}
Using Eqs.~(\ref{deltaT00}) and (\ref{dtt22}), 
 we can write down the solution for $\Psi$ as
\begin{eqnarray}
  &&\Psi(\eta,\chi,\bm \gamma)={1\over F(\eta)}\int_{\eta_*}^{\eta}d\eta_1
  {8\pi GF(\eta_1)\over B(\eta_1)}\delta T^0{}_{0(\phi)}(\eta_1,\chi ,\bm\gamma )
   \nonumber\\
   &&~~~~~~~~~~~~~\simeq -{1\over F(\eta)}\int_0^{\eta}d\eta_1
   {4\pi GF(\eta_1)\over B(\eta_1)}{m_0^2}\left(\phi(\eta_1,\chi,\bm \gamma)^2-  \phi(\eta_1,0)^2  \right)  ,
      \label{dtt3}
\end{eqnarray}
where the approximation $\eta_* / \eta \ll 1$, hence $\eta_* \simeq 0$ was used, and
we defined
\begin{eqnarray}
  &&F(\eta)=F_c\exp\left\{
    \int_0^\eta d\eta'{A(\eta')\over B(\eta')} \right\},
    \\
  &&A(\eta)=6{{\cal H}^2\over a^2}+24\pi G\rho_m
  ,
  ~~B(\eta)= 6{{\cal H}\over a^2},
\end{eqnarray}
{
with a constant $F_c$.
We note that the result of Eq.~(\ref{dtt3}) does not depend on the constant $F_c$. }
Under the condition $\Phi+\Psi=0$, Eq.~(\ref{abc}) becomes
\begin{eqnarray}
  &&{\Delta T\over T}(\bm\gamma)  
  \simeq2\int_{0}^{\eta_0} d\eta\left(
    {\partial \Psi(\eta,\chi,\bm \gamma) \over \partial \eta}
    \right)\biggl|_{\chi=\eta_0-\eta} .
    \label{dT/T}
\end{eqnarray}
Thus using Eq.~(\ref{dtt3}), 
 the two-point correlation function of temperature fluctuations from
the last scattering surface of the CMB is given by
\begin{eqnarray}
&&\biggl\langle{\Delta T\over T}(\bm\gamma) {\Delta T\over T}(\bm\gamma')\biggr\rangle 
\nonumber\\
&&\quad\quad
  = 4\int_{0}^{\eta_0} d\eta_1 \int_{0}^{\eta_0} d\eta_2
  \left[{\partial \over \partial \eta_1}{1\over F(\eta_1)}\int_0^{\eta_1}d\eta_3
    {4\pi G F(\eta_3)\over B(\eta_3)}m_0^2\right]
  \left[{\partial \over \partial \eta_2}{1\over F(\eta_2)}\int_0^{\eta_2}d\eta_4
    {4\pi G F(\eta_4)\over B(\eta_4)}m_0^2\right]
  \nonumber\\
  &&\quad\quad\quad\times\biggl\langle
  \left(\phi(\eta_3,\chi_3,\bm\gamma)^2- \phi(\eta_3,0)^2 
  % \phi(\eta'',\chi,\bm\gamma)^2|_{SA\chi=0}
   \right)
  \left(\phi(\eta_4,\chi_4,\bm\gamma')^2- \phi(\eta_4,0)^2 
  %\phi(\eta''',\chi',\bm\gamma')^2|_{SA\chi'=0}
  \right)
\biggr\rangle
  \biggl|_{\chi_3=\eta_0-\eta_1, \chi_4=\eta_0-\eta_2}.
\label{2pcf0}
\end{eqnarray}
The expectation value in (\ref{2pcf0}) can be decomposed into 
products of two-point functions by using the Wick-theorem in Eq.~(\ref{p2p2})
and calculated by using the two-point correlation function in Eq.~(\ref{2pcfphi}). 
The details of the calculation are given in Appendix \ref{App:2}, and we obtain
\begin{eqnarray}
&&\biggl\langle{\Delta T\over T}(\bm\gamma){\Delta T\over T}(\bm\gamma')\biggr\rangle
  =\int_{0}^{1} da_1 \int_{0}^{1} da_2
  \left[{\partial \over \partial a_1}{1\over F(a_1)}\int_0^{a_1}da_3
    {4\pi G m_0^2 F(a_3)\over 3a_3{H^2(a_3)}}\right]
  \left[{\partial \over \partial a_2}{1\over F(a_2)}\int_0^{a_2}da_4
    {4\pi G m_0^2 F(a_4)\over 3a_4{H^2}(a_4)}\right]
  \nonumber\\
  &&~~~~~~~~~~~~~~~~~~~~~~~~~~~\times(-4)
  \varphi^2(\eta_3)\varphi^2(\eta_4)
  \epsilon \left[-{2\over 3}R_1R_2
  \cos\psi-{2\over 15}
    R_1^2R_2^2\left({3\over 2}\cos^2\psi-{1\over 2}\right)\right],
\label{2pcf1}
\end{eqnarray}
where $R_1=\sqrt{-K}(\eta_0-\eta_1)$ and $R_2=\sqrt{-K}(\eta_0-\eta_2)$,
and $\eta_i$ are functions of $a_i$ as $\eta_i \equiv \eta(a_i)$ with $i=1,2,3,4$ respectively,
whose explicit form is given in Appendix \ref{appen:numer}, while $H(a)$ is
the Hubble parameter.

%\vspace{5mm}
In the rest of the paper, we will compare the result of Eq.~(\ref{2pcf1}) with the CMB observations
and constraint on the model parameters in the present scenario.
The multipole expansion of the angular two-point function of the CMB temperature fluctuation
is expressed as
\begin{eqnarray}
  \biggl\langle{\Delta T\over T}(\bm\gamma){\Delta T\over T}(\bm\gamma')\biggr\rangle
  ={1\over 4\pi}\sum_{\ell}(2\ell+1)C_\ell P_\ell(\cos\psi),
  \label{2pcfmul}
\end{eqnarray}
where $\cos\psi =\bm\gamma\cdot\bm\gamma'$.
Then, by comparing Eqs.~(\ref{2pcf1})~and~(\ref{2pcfmul}), it is explicit to find 
\begin{eqnarray}
&&  {3\over 4\pi}C_1 = S_1^2 {8\over 3}\epsilon \sim {\cal O}(\epsilon \Omega_K),
  \label{eq:c1}
  \\
 && {5\over 4\pi}C_2=S_2^2{8\over15}\epsilon\sim {\cal O}(\epsilon \Omega_K^2),
 \label{eq:c2}
\end{eqnarray}
where we define the coefficients $S_\ell$ by 
\begin{eqnarray}
 S_\ell=\int_{0}^{1} da \left(\sqrt{-K}(\eta_0-\eta(a))\right)^\ell
 {\partial \over \partial a}\left({1\over F(a)}\int_0^{a}da'
 {8\pi G \rho_{\rm DE}(a')F(a')\over 3{a'H^2(a')}}\right),
\label{abcd5}
\end{eqnarray}
where we used $\rho_{\rm DE}(a)=m_0^2\varphi^2/2$.
The approximate expression in the above formulae are obtained by 
$\sqrt{-K}\eta\sim\sqrt{-K}/H_0\sim \sqrt{\Omega_K}$.
We evaluate  higher multipoles in a similar manner, which are
approximately given by
\begin{eqnarray}
C_\ell \sim {\cal O}(\epsilon \Omega_K^\ell).
\end{eqnarray}
These higher multipoles with $\ell\geq3$ do not put tighter constraints compared with 
the dipole and the quadrupole as long as $\Omega_K\ll 1$. 
Thus, the dipole and the quadrupole
are the most important, which is reflected by the property that the typical scales
of the spatial variation are given by the supercurvature scale. 
Using the results for $S_\ell$ in Appendix \ref{appen:numer}, 
 numerical calculations of $S_\ell$ give the following results
\begin{eqnarray}
&& S_1 \simeq {1.1}\times10^{-1} \Omega_K^{1/2},
\label{coeff0}
\\
&& S_2 \simeq {0.9}\times10^{-1} \Omega_K,
\label{coeff}
\end{eqnarray}
where we assumed $\Omega_m=0.3$ and $\Omega_K \ll 1$.

The observed values of the dipole and the quadrupole in the CMB anisotropies
are found in the literature.
The dipole of the CMB is approximately expressed as
\begin{eqnarray}
{\delta T_{\rm{dipole}}\over T}={v\over c} \cos\theta,
\end{eqnarray}
where $v$ is the peculiar velocity of the observer
and $\cos\theta$ is the parameter related to the line-of-sight. The raw
observational result gives $v\approx {370} \rm{km/s}$~\cite{PeeblesTB,Planck2013xxvii}.
From this observation, we adopt the value of the dipole moment,
\begin{eqnarray}
  C_1^{\rm{obs}} \approx 6.3\times 10^{-6},
  \label{C1obs}
  \end{eqnarray}
where we used $3C_1/4\pi=(v/c)^2$. Comparing this with (\ref{eq:c1})~and~(\ref{coeff0}), we have the constraint from the dipole
\begin{eqnarray}
&&  \epsilon \Omega_K \simlt {4.9} \times 10^{-5}. 
\label{constr1}
\end{eqnarray}

The measurement of $C_2$ from the Planck Legacy Archive 
obtained with Planck satellite~\cite{Aghanim:2018eyx} with $1\sigma$ error is
\begin{eqnarray}
  {2\times3\over 2\pi}C_{2}^{\rm{obs}} =2.26^{+5.33}_{-1.32}\times
  {10^2\rm \mu K^2 \over (2.725 {\rm K})^2}. %\approx 1.0\times10^{-10}.
  \label{C2pobs}
  \end{eqnarray}
If we adopt the upper bound of the above observed value, taking the effect of the
observational error, we have
\begin{eqnarray}
  {2\times3\over 2\pi}C_{2}^{\rm{obs}} < 1.0\times10^{-10}.
  \label{C2pobs2}
  \end{eqnarray}
Then, Eq.~(\ref{C2pobs2}) with~(\ref{eq:c2}) and (\ref{coeff}) leads to
\begin{eqnarray}
\epsilon \Omega_K^2 < {1.0} \times 10^{-8}.
\label{constr2}
\end{eqnarray}

%%%%%%%%%%%%%%%%%%%%%%%%%%%%%%%%%%%%%%%%%%%%%%%%%%%%%%%%%%%%%%%%%%%%%%%%%%%%%%%%%%%%%%%%%%%%%%%%%%%%%%%%%%%%%%%%%%%%%%%%%%%%%%%%$
%%%%%%%%%%%%%%%%%%%%%%%%%%%%%%%%%%%%%%%%%%%%%%%%%%%%%%%%%%%%%%%%%%%%%%%%%%%%%%%%%%%%%%%%%%%%%%%%%%%%%%%%%%%%%%%%%%%%%%%%%%%%%%%%%
The constraints given by Eqs.~(\ref{constr1})~and~(\ref{constr2}) contain
the parameter $\epsilon$ describing some properties of 
the ancestor vacuum Eq.~(\ref{epsil}) along with the curvature parameter $\Omega_K$. 
The two parameters $\Omega_K$ and $\epsilon$ are coupled to each other in Eqs.~(\ref{constr1})
and (\ref{constr2}), which are natural outcomes because this scenario connects the spatial
curvature with the supercurvature-mode dark energy through the CDL tunneling inflation. 
Consequently, the constraint on the ancestor vacuum parameter $\epsilon$ is linked
with the value of the spatial curvature $\Omega_K$.
The upper bound of the spatial curvature is given by 
 $|\Omega_K|\simlt 10^{-2}\sim10^{-3}$ \cite{Planck2018}, and 
 if we take the possible value with BAO for $\Omega_K\sim10^{-3}$,
 the other parameter is constrained to satisfy the relation $\epsilon \simlt
 {10^{-2}}$.

%%%%%%%%%%%%%%%%%%%%%%%%%%%%%%%%%%%%%%%%%%%%%%%%%%%%%%%%%%%%%%%%%%%%%%%%%%%%%%%%%%%%%%%%%%%%%%%%%%%%%%%%%%%%%%%%%%%%%%%%%%%%%%%%$
%%%%%%%%%%%%%%%%%%%%%%%%%%%%%%%%%%%%%%%%%%%%%%%%%%%%%%%%%%%%%%%%%%%%%%%%%%%%%%%%%%%%%%%%%%%%%%%%%%%%%%%%%%%%%%%%%%%%%%%%%%%%%%%%%

\section{Conclusions}
\label{sec:sum}

We have studied a model of the dark energy in the universe
created by a bubble nucleation due to quantum tunneling from an ancestor vacuum. 
The supercurvature mode of an ultralight scalar field $\phi$
in the bubble of the present universe plays a role of the dark energy, which 
we call the supercurvature-mode dark energy.
In such a universe, 
the present universe is open and has a negative spatial curvature in the bubble
and fluctuations of the supercurvature modes are frozen on the
superhorizon scales. This is the reason that the mode behaves
as the dark energy in the present epoch.

In the present paper, we have particularly investigated 
large-scale inhomogeneity of the supercurvature-mode dark energy density. 
We show that 
the density contrast of the dark energy becomes of the order of one on the supercurvature scale $L_{sc}$,
which is much longer than the Hubble length $H_0^{-1}$ in the present universe; $L_{sc} \gg H_0^{-1}$,
and the spatial variation might be extremely tiny within $H_0^{-1}$.
Nevertheless, our calculations indicate that 
the large-scale inhomogeneity of the dark energy density can be detected
in the anisotropies of the CMB spectrum via the late-time ISW effect. 
The detectable signatures are imprinted at low angular momentum components
of the two-point correlation function of the CMB temperature fluctuation, especially
the dipole and the quadrupole. 
Comparing with the current observations of the CMB multipoles, 
we obtained upper bounds 
of the curvature parameter $\Omega_K$ and the ancestor vacuum parameter 
$\epsilon$, given in Eqs.~(\ref{constr1}) and (\ref{constr2}), respectively.
For example, if we assume that the spatial curvature is given by the current 
upper limit from observations, $\Omega_K\sim 10^{-3}$, 
the other parameter is given by $\epsilon\simlt 
10^{-2}$. 
For a smaller value of $\Omega_K$, $\epsilon$ can be larger.
Further investigations of the supercurvature-mode dark energy scenario
will be interesting in view of the large-scale CMB anomaly~(e.g.,~\cite{Rassat:2014yna,Erisken,Ade,Dominic,Aiola}).

String theory predicts axion-like particles(ALPs)~\cite{Arvanitaki,Witten} which are ultralight.
In an open inflation scenario created by a bubble nucleation of the true vacuum due to
quantum tunneling from the false ancestor vacuum,
the supercurvature-modes of these ultralight scalar fields provide a candidate for the dark energy. 

The supercurvature-mode dark energy scenario predicts a deviation of the equation of state
from the cosmological constant~\cite{Aoki} as well as the spatial variation presented in the present paper. 
The universe also predicts negative spatial curvature. 
Hence the model {could} be potentially verified / falsified by future observations.

%%%%%%%%%%%%%%%%%%%%%%%%%%%%%%%%%%%%%%%%%%%%%%%%%%%%%%%%%%%%%%%%%%%%%%%%%%%%%%%%%%%%%%%%%%%%%%%%%%%%%%%%%%%%%%%%%%%%%%%%%%%%%%%%$
%%%%%%%%%%%%%%%%%%%%%%%%%%%%%%%%%%%%%%%%%%%%%%%%%%%%%%%%%%%%%%%%%%%%%%%%%%%%%%%%%%%%%%%%%%%%%%%%%%%%%%%%%%%%%%%%%%%%%%%%%%%%%%%%%
\section*{Acknowledgments}

This work was supported by the Ministry of Education, Culture, Sports, Science and Technology (MEXT)/Japan Society
for the Promotion of Science (JSPS) KAKENHI Grant No.~15H05895, No.~16H03977
No.~17K05444, No.~17H06359 (KY), No.~17K14304 (DY), No.~16K05329, No.~18H03708 (SI).
We acknowledge Y. Sekino for collaboration at the early stage and also for 
critical comments. We also thank M. Sasaki for useful discussions.
%%%%%%%%%%%%%%%%%%%%%%%%%%%%%%%%%%%%%%%%%%%%%%%%%%%%%%%%%%%%%%%%%%%%%%%%%%%%%%%%%%%%%%%%%%%%%%%%%%%%%%%%%%%%%%%%%%%%%%%%%%%%%%%%$
%%%%%%%%%%%%%%%%%%%%%%%%%%%%%%%%%%%%%%%%%%%%%%%%%%%%%%%%%%%%%%%%%%%%%%%%%%%%%%%%%%%%%%%%%%%%%%%%%%%%%%%%%%%%%%%%%%%%%%%%%%%%%%%%$
\appendix

\section{Correlation function of the supercurvature mode}
\label{appen:scmde}

First, we recall the correlation function of the scalar field $\phi$ in the CDL geometry,
analytically continued to Lorentzian.
For more details, see Ref.~\cite{Aoki}. 
Taking only the contributions from the supercurvature modes with $k=i(1-\epsilon)$, 
it is given in Eq.~(4.5) in \cite{Aoki} as 
\begin{eqnarray}
 \langle \phi(\eta,R) \phi(\eta',0)\rangle^{(\rm{scm})}= \frac{-2 \pi i}{8 \pi^2 a(\eta)a(\eta')}
 \cdot {\rm Res}(i(1-\epsilon)) {\rm e}^{(1-\epsilon)
 (\eta+\eta'+2\tilde\eta_1)} \frac{1}{\sin \epsilon \pi}\frac{\sinh(1-\epsilon)R}{\sinh R},
 \label{appen:scm2pcf}
\end{eqnarray}
where $a(\eta)$ is the scale factor, and 
 ${\rm Res}(i(1-\epsilon))$ denotes the residue of the reflection coefficient ${\cal R}(k)$ 
 at the pole $k=i(1-\epsilon)$, whose explicit form is given in \cite{Aoki}. 
 $R$ is the radial coordinate parametrizing the spatial slice ${\rm H}^3$.
$\tilde\eta_1$
is a phase shift introduced for connecting the CDL and FLRW geometries smoothly, 
and can be expressed as
\begin{eqnarray}
\rm{e}^{\tilde\eta_1} = {\it H_A \over H_I}(1+\rm{e}^{2 X_0}),
\end{eqnarray}
where $X_0$ is related to the size of the bubble ($X_0 \rightarrow - \infty$ corresponds to a small bubble limit).
For small $\epsilon$, Eq.~(\ref{appen:scm2pcf}) reduces to Eq.~(\ref{2pcfphi}) with
\begin{eqnarray}
  && \varphi(\eta)=c_*^{1/2}{H_A^2\over m_A}
 \left({H_I\over H_A}\right)^\epsilon \varphi_*(\eta),
 \label{appen:varphi}
\end{eqnarray}
where $c_*$ is an ${\cal O}(1)$ constant (Eq.~(5.33) in \cite{Aoki}). 
$\varphi_*(\eta)$ represents the time evolution in the FLRW universe,
and, for instance, in the periods (ii) and (iii) in Sec. V-C of Ref.~\cite{Aoki}, it is given by
\begin{eqnarray}
 && \varphi_*(\eta)\simeq {\sin m_0t\over m_0t} ,
 \label{appen:phistar}
\end{eqnarray}
where $t$ is the proper time in the FLRW universe.
When $m_0 \simlt H_0$ is satisfied, $m_0 t \lesssim 1$, and we have $\varphi_*(\eta)\simeq 1$; the 
supercurvature mode is almost frozen. 

With the frozen supercurvature modes, we can set e.g. $\eta=0$. Then in the 
flat limit $\Omega_K\ll1$, the supercurvature modes behave as the dark energy 
with the density
\begin{eqnarray}
{8\pi G\over 3}\rho_{\rm DE}\simeq{8\pi G\over 3}{m_0^2\varphi^2(0)\over 2}=H_0^2
\Omega_{\Lambda}.
\end{eqnarray}
In the massless limit, $\epsilon \to 0$, and using a small bubble approximation $X_0 \rightarrow -\infty$, 
the well-known result for the coincident-point correlation function \cite{Linde}
\begin{eqnarray}
\langle \phi^2 \rangle={ \varphi^2(0) }=
{3 \over 8 \pi^2} {H_A^4 \over m_A^2}
\end{eqnarray}
is reproduced.

%%%%%%%%%%%%%%%%%%%%%%%%%%%%%%%%%%%%%%%%%%%%%%%%%%%%%%%%%%%%%%%%%%%%%%%%%%%%%%%%%%%%%%%%%%%%%%%%

\section{ Equation of state of dark energy}
\label{appen:eos}
%{\color{blue} 
 In the previous paper Ref.~\cite{Yamauchi}, the authors investigated the dynamical property 
  of the supercurvature-mode dark energy and showed that the EoS is modified
  from $w =-1.$ In this appendix, we show that this modification gives a higher order 
  correction to the large-scale spatial inhomogeneity and can be neglected in the 
  present investigation, which justifies our approximation using $w =-1.$
  %As was mentioned at the beginning of Section II, we justify our assumption of
  %EoS $w \simeq -1$ for the supercurvature-mode dark energy in this Appendix. 
  
  As was pointed out in Eq.~(3) in Ref.~\cite{Yamauchi} and Eq.~(5.59) in Ref.~\cite{Aoki},
  the contributions from the time-derivative terms to the pressure $p$ and the energy density $\rho$ 
  %of the energy momentum tensor $T^\mu{}_{\nu(\phi)}$ 
  %are given as ${\cal O}(\epsilon^2)$,and 
  can be ignored in comparison with the spatial-derivative terms
  % of the order ${\cal O}(\epsilon)$
  and mass terms as long as the conditions $m_0\ll H_0$ and $\epsilon \ll 1$ are satisfied.
  Then, the evolution of EoS $w$ of the supercurvature-mode dark energy yields
  \begin{eqnarray}
    w(z)=\frac{p}{\rho}=-\frac{1+\frac{2}{3}\tilde\epsilon (1+z)^2}{1+2\tilde\epsilon (1+z)^2}=-1+{\frac{4}{3}\tilde\epsilon(1+z)^2 \over 1+ 2\tilde\epsilon(1+z)^2},
    \label{eos1}
  \end{eqnarray}  
  where $z$ is the cosmological redshift and $\tilde\epsilon$ is defined as
  \begin{eqnarray}
    \tilde\epsilon\equiv \frac{1}{(m_0 / H_0)^2} \epsilon \Omega_K \simgt \mathcal{O}(\epsilon\Omega_K),
    \label{eq:tilep}
  \end{eqnarray}  
  with $\Omega_K=1/(H_0^2 R_c^2)$. In the above approximate equality, 
  we assumed $m_0\lesssim H_0$; e.g., $ 10^{-1} H_0 \lesssim m_0 \lesssim H_0$.
  %$m_0\ll H_0$ y in practice; instead, will serve for our interest here, 
  %although this depends on the choice of the parameter $m_0$.
 % For example, for $ 10^{-1} H_0 \lesssim m_0 \lesssim H_0$, 
  Then, under the condition of Eq.~(\ref{constr1}), $\tilde\epsilon \ll 1$ follows.
  %\begin{eqnarray}
  %  \tilde\epsilon\equiv \frac{1}{(m_0 H_0)^2} \epsilon \Omega_K,
  %  \label{eq:tilep}
  %\end{eqnarray} 
  If we define the present EoS and its derivative by $w_0\equiv w(z=0)$ 
  and $w_1 \equiv -a \frac{dw}{da}|_{a=1}$ in the 
  Chevallier-Polarski-Linder (CPL) parametrization~\cite{ChePolar,Linde0}, 
   Eq.~(\ref{eos1}) gives the deviation of EoS $w_0$ from $-1$ as
  \begin{eqnarray}
    w_0+1={\frac{4}{3}\tilde\epsilon \over 1+ 2\tilde\epsilon}\simeq \frac{4}{3}\tilde\epsilon,
    \label{eosdev}
  \end{eqnarray}  
 and its derivative as
  \begin{eqnarray}
    w_1={\frac{8}{3}\tilde\epsilon \over (1+ 2\tilde\epsilon)^2} \simeq \frac{8}{3}\tilde\epsilon=2(w_0+1).
    \label{eosderiv}
  \end{eqnarray}  
  Eqs.~(\ref{eosdev})~and~(\ref{eosderiv}) confirm that the deviation of dark energy EoS from $w=-1$ at late times 
  before and at present epoch is small as long as $\tilde \epsilon \ll 1$ holds. The deviation from $w=-1$ gives a higher order
  correction to the investigation of large-scale inhomogeneity and can be neglected 
  in the leading order calculations of the $C_l$'s.
%}

%%%%%%%%%%%%%%%%%%%%%%%%%%%%%%%%%%%%%%%%%%%%%%%%%%%%%%%%%%%%%%%%%%%%%%%%%%%%%%%%%%%%%%%%%%%%%%%%
\section{One-point probability function of dark energy}
\label{appen:pdf2}
\begin{figure}
\includegraphics[width=0.45\textwidth]{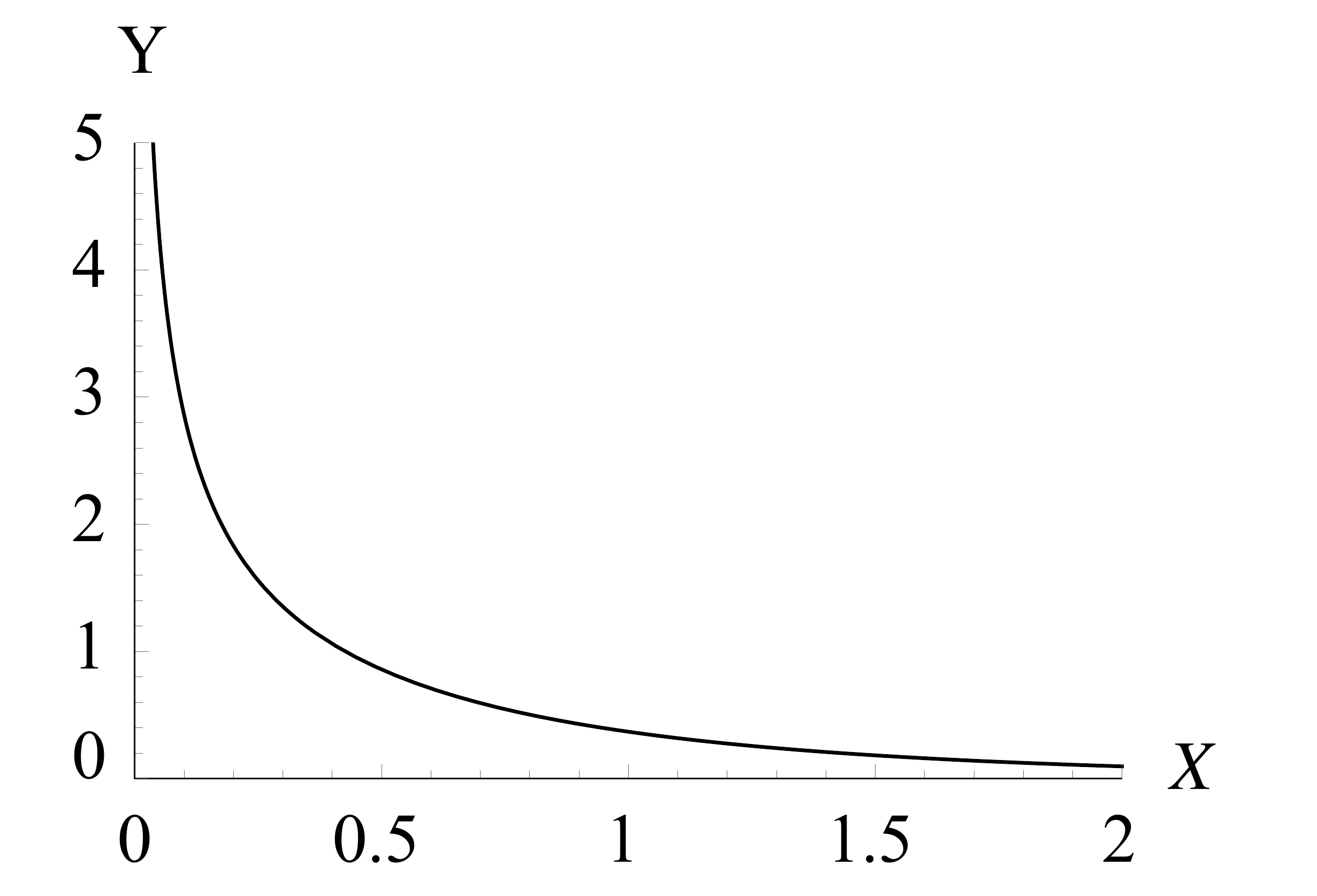} %eps}
\caption{Probability density function $f(\rho_{\rm DE})$ as a function of $\rho_{\rm DE}$.
  The horizontal axis is $X=\rho_{\rm DE}/m_0^2\varphi^2(0)$, and the vertical
  axis is $Y=\sqrt{4\pi}m_0^2\varphi^2(0)f(\rho_{\rm DE})$.
\label{fig-pdf_omega_lambda2}}
\includegraphics[width=0.45\textwidth]{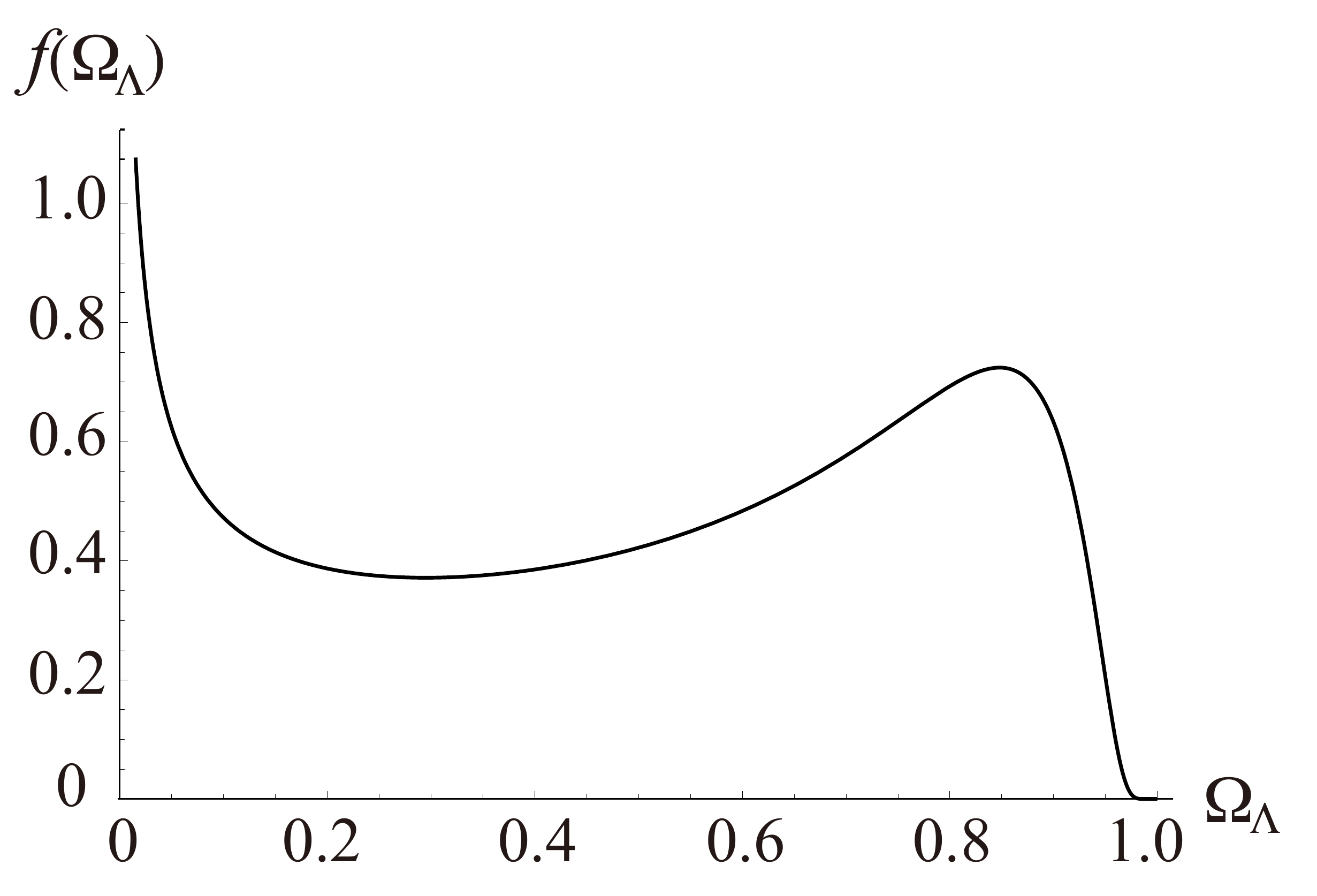} %eps}
\caption{Probability density function $f(\Omega_{\Lambda})$ of $\Omega_{\Lambda}$ with
  its expectation value fixed as $\bar{\Omega_{\Lambda}}=0.7$.}
\label{fig:pdf_omega_lambda}
\end{figure}

In this Appendix, we demonstrate the explicit form of the probability functions
of the dark energy density and the density parameter.
For a {\it normalized} probability variable of the field, 
the distribution function is given by
\begin{eqnarray}
  P(\widetilde\phi(\bm x))={1\over \sqrt{2\pi}}\exp\left[-{1\over 2}{\widetilde\phi^2(\bm x)}\right]. 
\end{eqnarray}
We note that $\langle \widetilde\phi^2(\bm x)\rangle=1$.
Using $\widetilde\phi(\bm x)$, we may write the scalar field as
$\phi(\eta ,\bm x)=\varphi (\eta )\,\widetilde\phi(\bm x)$,
where $\varphi(0)$ is defined in Appendix \ref{appen:scmde}.
We find the probability density function of the supercurvature-mode dark energy
density, given by
%\begin{eqnarray}
\begin{eqnarray}
  \rho_{\rm DE}( \bm x)={1\over2}m_0^2 \phi^2(\eta_0, \bm x)\approx
      {1\over 2}m_0^2\varphi^2(0) \widetilde\phi^2(\bm x).
      \label{rhodea}
\end{eqnarray}
On the large scales $R > R_{sc}$, the spatial variation is significant, however,
as long as we consider a region of the present Hubble horizon,
which is much smaller than the scale $R_{sc}$, $\rho_{\rm DE}( \bm x)$ can be regarded
as a probability variable through $\widetilde\phi$ by Eq.~(\ref{rhodea}).
Following the conservation of the probability, 
\begin{eqnarray}
d\widetilde\phi(\bm x)P(\widetilde\phi(\bm x))=d \rho_{\rm DE}\,
f(\rho_{\rm DE}),
\end{eqnarray}
we define the probability density function of $\rho_{\rm DE}(\bm x)$
\begin{eqnarray}
  f(\rho_{\rm DE})=\int d\widetilde\phi(\bm x) \delta(\rho_{\rm DE}-\rho_{\rm DE}(\bm x))P(\widetilde\phi(\bm x)). 
\end{eqnarray}
It can be analytically calculated as
\begin{eqnarray}
  f(\rho_{\rm DE})={1\over \sqrt{4\pi} m_0^2\varphi^2(0)}{\exp\left(-{\rho_{\rm DE}/m_0^2\varphi^2(0)}\right)\over 
  \sqrt{\rho_{\rm DE}/ m_0^2\varphi^2(0)}},
\end{eqnarray}
which is plotted in Figure~\ref{fig-pdf_omega_lambda2}.
This figure demonstrates a wide distribution of probability of $\rho_{\rm DE}$
at scales larger than the supercurvature scale $R_{sc}$
even when we fix the parameter as Eq.~(\ref{expectationrhode}). 

We also discuss the probability density function of the dark energy density parameter
defined by 
\begin{eqnarray}
  \Omega_\Lambda({\bm x}%\widetilde\phi
) = \frac{\rho_{\rm DE}{(\bm x)}}{\rho_{\rm DE}{(\bm x)}+\rho_m} =
  \frac{\bar\Omega_\Lambda\widetilde\phi^2{(\bm x)}}
       {1-\bar\Omega_\Lambda+\bar\Omega_\Lambda\widetilde\phi^2{(\bm x)}},
\end{eqnarray}
where $\rho_m$ is the dark matter energy density.
In a similar way to the case for the dark energy density,
we can find the probability density function of $\Omega_\Lambda$ as
\begin{eqnarray}
  f(\Omega_\Lambda)=\int d\widetilde\phi \delta(\Omega_\Lambda-\Omega_\Lambda(
  {\bm x}%\widetilde \phi
)) P(\widetilde\phi{(\bm x)}). 
\end{eqnarray}
It can be analytically calculated as
\begin{eqnarray}
f(\Omega_\Lambda)&=&\frac{1}{2\sqrt{2\pi}\Omega_\Lambda(1-\Omega_\Lambda)}
\sqrt{\frac{\Omega_\Lambda(1-\bar\Omega_\Lambda)}{\bar\Omega_\Lambda(1-\Omega_\Lambda)}}
\exp\left(-\frac{\Omega_\Lambda(1-\bar\Omega_\Lambda)}{2\bar\Omega_\Lambda(1-\Omega_\Lambda)}\right).
\label{fol}
\end{eqnarray} 
Figure~\ref{fig:pdf_omega_lambda} plots the function $f(\Omega_\Lambda)$ assuming
$\bar \Omega_\Lambda=0.7$ in Eq.~(\ref{fol}). 
$f(\Omega_\Lambda)$ has a peak at a point of $\Omega_\Lambda$ slightly larger
than $\bar\Omega_\Lambda=0.7$, but this figure demonstrates a wide distribution
of probability of $\Omega_\Lambda$ at scales larger than the supercurvature scale $R_{sc}$.

%%%%%%%%%%%%%%%%%%%%%%%%%%%%%%%%%%%%%%%%%%%%%%%%%%%%%%%%%%%%%%%%%%%%%%%%%%%%%%%%%%%%%%%%%%%%%%%%
\section{Derivation of Eq.~(\ref{2pcf1})} \label{App:2}
The expectation value in (\ref{2pcf0}) can be decomposed into 
products of two-point functions by using the Wick-theorem in Eq.~(\ref{p2p2}):
\begin{eqnarray}
  \langle(\phi^2(X)-\phi^2(0))(\phi^2(X')-\phi^2(0'))\rangle
  =2\left(\langle\phi(X)\phi(X')\rangle^2
  -\langle\phi(X)\phi(0')\rangle^2
  -\langle\phi(0)\phi(X')\rangle^2
  +\langle\phi(0)\phi(0') \rangle^2\right) .
  \label{pppp}
\end{eqnarray}
Here, $X, X', 0, 0'$ denote $(\eta,\chi,\bm \gamma)$, $(\eta',\chi',\bm \gamma')$, 
$(\eta, 0, \gamma)$, and $(\eta', 0, \gamma')$, respectively.
Then, using the two-point correlation function given in Eq.~(\ref{2pcfphi}), Eq.~(\ref{pppp}) 
can be evaluated as 
\begin{eqnarray}
  &&\langle(\phi^2(X)-\phi^2(0))(\phi^2(X')-\phi^2(0'))\rangle
  \nonumber\\
  &&~~~~~~~~
= 2\varphi^2(\eta)\varphi^2(\eta')\left(
          {\sinh^2(1-\epsilon)R\over(1-\epsilon)^2\sinh^2 R}
          -{\sinh^2(1-\epsilon)R_1\over(1-\epsilon)^2\sinh^2 R_1}
          -{\sinh^2(1-\epsilon)R_2\over(1-\epsilon)^2\sinh^2 R_2}+1\right)
          \nonumber\\
          &&  ~~~~~~~~
=-4\varphi^2(\eta)\varphi^2(\eta')\left(
          R\coth R-R_1\coth R_1-R_2\coth R_2+1
          \right)\epsilon
          +{\cal O}(\epsilon^2)
          \nonumber
          \\
          &&          ~~~~~~~~
          \simeq-4\varphi^2(\eta)\varphi^2(\eta')\left(
          \frac{1}{3} \left(R^2-R_1^2-R_2^2\right)+\frac{1}{45}
          \left(-R^4+R_1^4+R_2^4\right)+{\cal O}\left(R^6\right)\right)\epsilon, 
  \label{pppp2}
\end{eqnarray}
where  $ R_i= \sqrt{-K}\chi_i$ for $i=1,2$. 
The schematic relation of $R$, $R_1$, and $R_2$ is presented in Fig.~\ref{fig:schematic}. 
In the expansion of $\coth(R_i)$, we used $R_1=\sqrt{-K}\chi_1\ll1$ and $R_2=\sqrt{-K}\chi_2\ll1$.

Using the relation of Eq.~(\ref{RRRR}), we have
\begin{eqnarray}
    \frac{1}{3} \left(R^2-R_1^2-R_2^2\right)+\frac{1}{45}
    \left(-R^4+R_1^4+R_2^4\right)
    &&\simeq-{2\over 3}R_1R_2\left(1-{2\over 15}\left(R_1^2+R_2^2\right)\right)
    \cos\psi-{2\over 15}
    R_1^2R_2^2\left({3\over 2}\cos^2\psi-{1\over 2}\right)
    \nonumber
    \\
    &&\simeq-{2\over 3}R_1R_2\cos\psi-{2\over 15}
    R_1^2R_2^2\left({3\over 2}\cos^2\psi-{1\over 2}\right).
    \label{approxR}
\end{eqnarray}
Substituting Eqs.~(\ref{pppp2})~and~(\ref{approxR}) into Eq.~(\ref{2pcf0}), we obtain 
Eq.~(\ref{2pcf1}).

%%%%%%%%%%%%%%%%%%%%%%%%%%%%%%%%%%%
\section{Estimations of $S_\ell$}
\label{appen:numer}
The conformal time $\eta$ and scale factor $a$ are related by 
\begin{eqnarray}
  {1\over a^2}{d a\over d\eta}=H(a),
\label{aaaH}
\end{eqnarray}
where the evolution of the Hubble parameter obeys the Friedmann equation. When the dark energy
is approximated by the cosmological constant, we may express
\begin{eqnarray}
  H^2(a)={8\pi G\over 3}\left(\rho_{m}+\rho_{\rm DE}\right)-{K\over a^2}
  \equiv H_0^2\biggl({\Omega_m \over a^3}+{\Omega_K\over a^2}+(1-\Omega_m-\Omega_K)\biggr).
\end{eqnarray}
We assume a nearly flat FLRW universe by adopting
$\Omega_m\approx0.3$, $\Omega_\Lambda\approx 0.7$,
and $\Omega_K\approx0$, and we approximate $S_\ell$ defined by Eq.~(\ref{abcd5}) as
\begin{eqnarray}
S_\ell=\int_{0}^{1} da \left(\sqrt{-K}(\eta_0-\eta(a))\right)^\ell
{\partial \over \partial a}\left({G(a)\over F(a)}\right),
\label{EqFga}
\end{eqnarray}
where
\begin{eqnarray}
  &&G(a)=\int_0^{a}da'   {8\pi G \rho_{\rm DE}(a')F(a')\over 3{a'H^2(a')}}=
  \int_0^{a}da'   {(1-\Omega_m)a'{}^2F(a')\over \Omega_m+(1-\Omega_m)a'{}^3},
  %\simeq
  %\int_0^a da' {1-\Omega_m\over \Omega_m/a'^3+1-\Omega_m}.
  \label{defFga}  
  \\
  &&F(a)=F_c\exp\left\{\int_0^{a}{da'\over a'}\left(1+{3\Omega_m\over 2[\Omega_m+(1-\Omega_m)a'{}^3]}\right)\right\}
    {\equiv{F_ca^{5/2}\over \sqrt{\Omega_m+(1-\Omega_m)a^3}}}.
    \label{defFa}
\end{eqnarray}
From Eq.~(\ref{aaaH}), the conformal time is written as
\begin{eqnarray}
  &&\eta(a)=\int^{a}_0\frac{da'}{a{'}^2H(a')}=H_0^{-1}\int^a_0\frac{da'}{a'^2(1-\Omega_m+\Omega_m a'^{-3})^{1/2}} .
\label{etaa}
\end{eqnarray}
An approximate expression for $S_\ell$ is given by
substituting (\ref{defFga}), (\ref{defFa}) and (\ref{etaa}) into (\ref{EqFga}).
We evaluate numerically the integrations over $a$,
and obtain Eqs.~(\ref{coeff0})~and~(\ref{coeff}).

%%%%%%%%%%%%%%%%%%%%%%%%%%%%%%%%%%%%%%%%%%%%%%%%%%%%%%%%%%%%%%%%%%%%%%%%%%%%%%%%%%%%%%%%%%%%%%%%%%%%%%%%%%%%%%%%%%%%%%%%%%%%%%%%%%%%%%

\end{document}